\documentclass{article}

\usepackage{PRIMEarxiv}

\usepackage[utf8]{inputenc} 
\usepackage[T1]{fontenc}    
\usepackage{hyperref}       
\usepackage{url}            
\usepackage{booktabs}       
\usepackage{amsfonts}       
\usepackage{nicefrac}       
\usepackage{microtype}      
\usepackage{fancyhdr}       
\usepackage{graphicx}       
\graphicspath{{media/}}     
\usepackage{xcolor}
\usepackage{float}
\usepackage{multirow}
\usepackage{array}
\usepackage{setspace}
\usepackage{caption}
\usepackage{subcaption}
\usepackage{amssymb}
\usepackage{amsmath}
\usepackage{soul}

\newcolumntype{L}[1]{>{\raggedright\arraybackslash}p{#1}}

\title{Demystifying the Design Space and Best Practices for Heterogeneous LLM Inference and Serving}
\author{
    \textbf{Zhixin Wang\textsuperscript{1}, Zhengbo Wang\textsuperscript{4,6}, Fangcheng Fu\textsuperscript{2,}\thanks{Corresponding author. ccchengff@sjtu.edu.cn},  Yinhui Lu\textsuperscript{3}, Jinlong Hou\textsuperscript{1}, Yijie Chen\textsuperscript{5}} \\
    \textbf{Xiaowei Shen\textsuperscript{5}, He Liu\textsuperscript{6}, Xiangbin Li\textsuperscript{6}, Jun Chen\textsuperscript{6}, Ruya Gu\textsuperscript{6}, Dian Wang\textsuperscript{5}, Zhou Tan\textsuperscript{5}} \\
    \textbf{Yuan Cheng\textsuperscript{1}, Hongzhou Zhang\textsuperscript{7}, Xiangjun Huang\textsuperscript{5}, Ping Zhang\textsuperscript{6}, Xiaohe Hu\textsuperscript{1, 6}} \\
    \\
    \textsuperscript{1}Shanghai Innovation Institute \\
    \textsuperscript{2}Shanghai Jiao Tong University \\
    \textsuperscript{3}Fudan University \\
    \textsuperscript{4}Renmin University of China \\
    \textsuperscript{5}MetaX Integrated Circuits (Shanghai) Co., Ltd. \\
    \textsuperscript{6}Infrawaves \\
    \textsuperscript{7}Shanghai AI Power Technology Co., Ltd.
}

\begin{document}

\maketitle

\begin{abstract}
Heterogeneous prefill-decode (PD) inference is now in production: prefill on cost-efficient or supply-available accelerators, decode on bandwidth-strong ones, and KV state crossing mixed interconnects in mixed numerical formats. Each deployment makes these decisions on its own. What is missing is the picture across configurations—which decisions must be made jointly at the PD boundary, and which can be made independently.
We propose a design space organized along four design axes—accelerator, precision, interconnect, and KV residency—and the workload regime (stage pressure) they respond to. We show that only a subset of interactions among these factors become binding constraints once PD inference becomes heterogeneous. These interactions surface through three recurring boundary decisions: compute placement, KV representation, and KV ownership.
The resulting analysis yields concrete guidance. Precision policy belongs to runtime roles rather than to a single system-wide setting, because the same low-bit format relieves different bottlenecks on each side of the boundary. KV transfer engines move bytes rather than tensor semantics, making representation compatibility an explicit boundary concern whenever producer and consumer differ. The KV handoff also carries a lifecycle—reservation, release, and failure recovery—that spans prefill and decode and requires explicit ownership.
Two further interactions remain open. Cross-vendor and interconnect-related claims are stated as design guidance grounded in industrial deployment observations and source-code inspection of the runtimes involved.
\end{abstract}

\section{Introduction}

Large language model (LLM) serving is becoming heterogeneous by necessity rather than design preference. Production deployments increasingly combine accelerators with different cost profiles, memory capacities, software stacks, and availability constraints. At the same time, serving systems have adopted prefill-decode (PD) disaggregation~\cite{zhong_distserve_2024,patel_splitwise_2023}, low-precision execution~\cite{lin_awq_2024,lin_qserve_2024}, Key-Value (KV) cache offloading~\cite{liu_lmcache_2025}, and distributed KV transfer~\cite{qin_mooncake_2024} to improve efficiency and scalability. As a result, a single request may be prefilled on one accelerator, decoded on another, transferred through a separate communication substrate, and served with KV state that resides outside GPU memory.

Most prior work studies these mechanisms independently. PD-disaggregated serving separates prompt processing from autoregressive generation. Quantization reduces compute and memory cost through lower-precision weights, activations, and KV caches. KV-centric systems introduce transfer engines and multi-tier KV storage. Heterogeneous serving platforms place different stages on different accelerators. Each line of work optimizes one dimension of the serving stack. In production deployments, however, these dimensions increasingly co-exist.

Once prefill and decode run on different resources, choices that seem independent can start affecting each other. A precision policy for quantization that works well on one accelerator may have no executable kernel path on another. A KV representation produced by one runtime may not be directly consumable by another, diminishing the effectiveness of KV-centric optimizations. A residency strategy that is feasible under a same-vendor GPU-direct path may become latency-prohibitive when the deployment falls back to CPU-aided transfer. In other words, the prefill-decode boundary turns many local configuration choices into decisions that must be coordinated across stages.

A concrete example illustrates the issue. Consider a PD-disaggregated deployment in which prefill and decode use different KV-cache formats. Existing KV transfer engines such as NIXL~\cite{vllm_nixl_source_2026} and Mooncake move buffers as raw bytes and generally do not negotiate tensor semantics such as dtype, quantization metadata, or layout. If producer and consumer disagree on the representation, the transfer path itself cannot detect the mismatch. The result is not a transport failure but a correctness failure: bytes arrive successfully while the receiver interprets them under a different numerical contract. The problem is not caused by quantization alone, nor by the interconnect alone, but by the interaction between them at the PD boundary.

This observation motivates the central question of this paper: \textbf{When heterogeneous PD inference combines different accelerators, numerical formats, interconnect paths, and KV residency tiers, which design decisions must be made jointly and which can be made independently?}

We study heterogeneous PD inference as a design space defined by four design axes—accelerator, precision, interconnect, and KV residency—and the workload regime that drives their selection. Table~\ref{tab:axes} summarizes these decision variables. Rather than treating these dimensions as independent tuning knobs, we focus on the couplings that emerge at the prefill-decode boundary. Our key observation is that only a subset of cross-axis relationships become binding system constraints. These couplings expose three boundary decisions that every heterogeneous deployment must eventually resolve:

\begin{itemize}
\item 
Compute placement: which accelerator serves each stage, which precision paths are executable on that accelerator, and how workload pressure shapes the stage-to-resource mapping.

\item 
KV representation: how Runtime KV State is represented, transferred, translated, and consumed across heterogeneous runtimes and interconnects.

\item 
KV ownership and lifecycle: where KV capacity is reserved, when it is released, and how failures, cancellation, and congestion are handled across the handoff.
\end{itemize}

These decisions appear repeatedly across existing serving frameworks, despite differences in implementation. They surface in accelerator assignment, precision selection, KV-cache quantization, transfer-engine design, cache residency management, and handoff protocols. Viewing them as boundary decisions reveals dependencies that are difficult to see when each mechanism is studied in isolation.

The contribution of this paper is therefore not a new serving or inference architecture, scheduling algorithm, or quantization method. Instead, we provide a design-space view of heterogeneous PD inference and derive a set of deployment-oriented best practices. We identify the tightest couplings among accelerator, precision, interconnect, KV residency, and workload regime; organize them around the three boundary decisions above; and use a combination of controlled measurements, source-code inspection, and industrial deployment observations to illustrate their consequences.

\begin{table}[!b]
\centering
\caption{Five design-space axes for heterogeneous PD inference decisions.}
\label{tab:axes}
\small
\begingroup
\renewcommand{\arraystretch}{1.28}
\begin{tabular}{@{}L{0.10\linewidth}L{0.18\linewidth}L{0.66\linewidth}@{}}
\toprule
Axis & Name & What it describes \\
\midrule
A1 & Accelerator & Hardware backend serving a stage, including compute capability, memory system, kernel ecosystem, cost, and availability. \\
\addlinespace[2pt]
A2 & Precision & Numerical representation of weights, activations, and KV state, including format, layout, and conversion behavior. \\
\addlinespace[2pt]
A3 & Interconnect & Communication path used to move Runtime KV State across the PD boundary. \\
\addlinespace[2pt]
A4 & Stage Pressure & Workload-driven pressure on each stage, including prompt length, generation length, prefix reuse, and latency objectives. \\
\addlinespace[2pt]
A5 & KV Residency & Placement of Runtime KV State across memory and storage tiers. \\
\bottomrule
\end{tabular}
\endgroup
\end{table}

The resulting framework provides a structured way to reason about heterogeneous inference systems and a practical guide for deciding which choices can remain local and which require explicit coordination across the prefill-decode boundary.

\begin{figure}[t]
\centering
\includegraphics[width=0.84\linewidth]{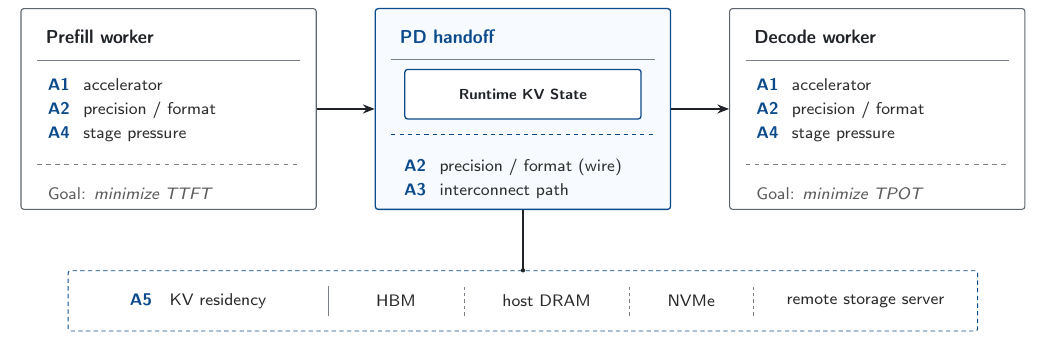}
\caption{Five-axis view of heterogeneous PD inference around Runtime KV State.
Prefill produces the state, the PD handoff transfers it, and decode consumes it
under accelerator, precision/format, interconnect, stage-pressure, and KV
residency constraints.}
\label{fig:design-space}
\end{figure}

\section{The PD Boundary}
\label{sec:design-space}

Prefill-decode (PD) disaggregation separates prompt processing from autoregressive generation, but the separation is more than a scheduling decision. Once the two stages run on different resources, the system must transfer state from one stage to the other. In homogeneous deployments, this transfer is often treated as an implementation detail because both stages share the same accelerator stack, numerical assumptions, and memory model. In heterogeneous deployments, the handoff itself becomes a first-class design problem.

The central object of this paper is the Runtime KV State: the information produced by prefill and required by decode to continue execution. Runtime KV State includes not only KV tensors, but also the representation, metadata, residency information, and ownership state needed for successful consumption. Figure~\ref{fig:design-space} illustrates this perspective. Prefill produces Runtime KV State, the state crosses a handoff boundary, and decode consumes it. Along that path, the state may traverse different accelerators, numerical formats, transfer mechanisms, and storage tiers.

Viewing heterogeneous serving through Runtime KV State provides a common abstraction across existing systems. Whether a deployment uses NIXL, Mooncake, LMCache, host-staged transfer, or direct GPU-to-GPU communication, the same fundamental question remains: what assumptions made by the producer must still hold when the state reaches the consumer?

\subsection{Design Axes Behind the Boundary}

We characterize heterogeneous PD inference using four design axes and one workload dimension. These factors are routinely tuned in production systems. Accelerators are chosen according to performance and availability. Precision is adjusted to trade quality for efficiency. Different interconnect paths provide different bandwidth and latency characteristics~\cite{licker_fabriclib_2025,liao_ubmesh_2025}. KV state may remain in GPU memory, move to host memory, or be retained in a shared cache. Meanwhile, workload characteristics continuously change the pressure placed on prefill and decode resources.

If these dimensions were independent, heterogeneous serving would be largely a configuration problem. In practice, they are not. Decisions made along one axis frequently constrain what is feasible along another. A precision format may be valid on one accelerator but not another. A residency strategy may be practical under direct RDMA but not under a host-staged path. A transfer engine may successfully move bytes while remaining unaware that producer and consumer disagree on the representation being transferred.
The challenge is therefore not understanding each axis individually, but understanding where interactions among them become system-level constraints. 

\begin{figure}[t]
\centering
\includegraphics[width=0.88\linewidth]{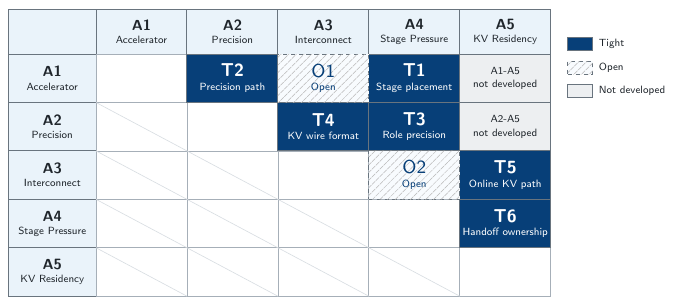}
\caption{Coupling matrix over the five design-space axes for heterogeneous PD
inference. Upper-triangular cells mark Tight couplings (T1--T6), Open
couplings (O1--O2), and pairs not developed separately.}
\label{fig:couplings}
\end{figure}

\subsection{Three Boundary Decisions}

We organize heterogeneous PD inference around four design axes—accelerator, precision, interconnect, and KV residency—together with workload-induced stage pressure. Figure~\ref{fig:couplings}
summarizes the pairwise interactions among these axes, highlighting a subset of tight couplings that become binding under disaggregated execution. These couplings can be interpreted through three boundary decisions: compute placement, KV representation, and KV ownership.

\textbf{Compute Placement (\S\ref{sec:placement}).} The first decision determines where computation executes. A deployment must decide which accelerator pool serves prefill, which serves decode (T1), and which precision paths are executable on those resources (T2). The decision is shaped by workload pressure because prefill and decode expose different bottlenecks. Those bottlenecks also determine where a precision path pays off: compute-oriented formats may matter most in prefill, while lower-bit decode paths can relieve memory and KV pressure (T3). Long prompts, long generations, prefix reuse, and latency objectives all influence the desired stage-to-resource mapping.

\textbf{KV Representation (\S\ref{sec:representation}).} The second decision determines how Runtime KV State is represented (T4) and transferred (T5) across the boundary. This includes the numerical format of the KV state, the representation visible to the transfer path, and the format expected by the decode runtime. In homogeneous deployments these representations often coincide. In heterogeneous deployments they may diverge, making translation, validation, and ownership of conversion explicit design concerns.

\textbf{Ownership and Lifecycle (\S\ref{sec:handoff}).} The third decision determines how Runtime KV State is managed after it is produced (T6). A deployment must decide where capacity is reserved, which component owns retained state, when resources are released, and how cancellation, timeout, and failure recovery are handled. These questions become increasingly important when decode admission, cache residency, and transfer timing vary under changing workload conditions.

Together, these three decisions provide a deployment-oriented view of heterogeneous PD inference. Rather than treating accelerator, precision, interconnect, workload, and residency as independent tuning knobs, we view them as inputs to a smaller set of boundary decisions that every heterogeneous deployment must eventually resolve.

The remainder of the paper develops these decisions through a coupling analysis. We identify the tightest cross-axis dependencies, organize them according to the boundary decision they influence, and derive deployment-oriented best practices for each.

\section{Matching Resources to Stage Demand} 
\label{sec:placement}

\subsection{Why This Decision Exists} 

The primary motivation for PD disaggregation is that prefill and decode place fundamentally different demands on the serving system~\cite{hu_tetriinfer_2024}. Prefill processes large batches of prompt tokens and typically benefits from high compute throughput. Decode performs autoregressive token generation by iteratively attending over an expanding KV cache, and is typically constrained by memory bandwidth, KV locality, and latency requirements. As workload characteristics change, the relative pressure on the two stages changes as well.

In homogeneous deployments, both stages are typically executed on the same hardware platform, making stage placement largely a scheduling problem. Heterogeneous deployments remove this assumption. Once multiple accelerator types and precision paths become available, operators must decide not only how much capacity to allocate to each stage, but also which resources are best suited for serving that stage.
Consequently, a hardware platform that performs well under one workload regime may become inefficient under another, and a precision format that is theoretically supported may not provide a practical execution path if kernel coverage or runtime support is limited.

\begin{figure}[t]
\centering
\includegraphics[width=0.9\linewidth]{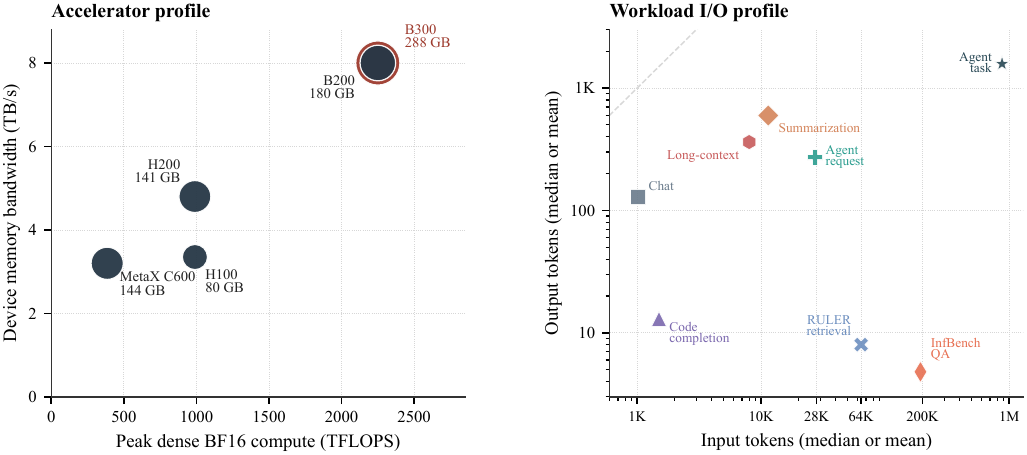}
\caption{Accelerator and workload profiles motivating accelerator-stage
matching. Accelerator points show peak dense BF16/FP16 compute, memory bandwidth,
and HBM capacity~\cite{nvidia_h100_specs_2026,nvidia_h200_specs_2026,nvidia_dgx_b200_specs_2026,nvidia_hgx_platform_specs_2026,nvidia_blackwell_ultra_datasheet_2026,metax_c600_specs_2026}; workload points show representative input-output token
regimes~\cite{patel_splitwise_2023,huang_efficient_attentions_2021,qin_mooncake_2024,ding_mcp_agents_2025,hsieh_ruler_2024,zhang_infinitebench_2024}.}
\label{fig:accel-landscape}
\end{figure}

This decision is more complex than matching the fastest hardware to the busiest stage. Effective placement depends on the interaction among workload pressure, accelerator capabilities, and executable precision paths. Figure~\ref{fig:accel-landscape} places representative accelerator and workload profiles next to this decision.

\subsection{Critical Dependencies}

Three dependencies consistently influence compute placement decisions.

\textbf{T1: Accelerator $\leftrightarrow$ Stage Pressure.} Different stages stress different resources. Long prompts and large batches increase compute demand during prefill, while long generations and large active-session counts increase pressure on decode memory bandwidth and KV capacity. As workload characteristics shift, the preferred stage-to-resource mapping may shift as well. This pressure therefore determines the P:D resource balance in a PD-disaggregated deployment, requiring the system to allocate prefill and decode resources in proportion to workload-induced QPS demand.

\textbf{T2: Accelerator $\leftrightarrow$ Precision.} Precision support is ultimately determined by executable software paths rather than advertised hardware features~\cite{lu_mixquant_2026}. Two accelerators may nominally support the same numerical format while exposing very different kernel maturity, scheduling behavior, or operational stability. Placement decisions must therefore be based on achievable performance rather than precision support in isolation.

\textbf{T3: Precision $\leftrightarrow$ Stage Pressure.} The value of a precision format depends on the stage in which it is used. During prefill, higher-throughput formats may improve dense compute efficiency~\cite{chen_pmpd_2024,zhang_qqq_2024} with limited impact on overall memory footprint. During decode, lower-bit representations can directly reduce memory traffic and KV-related pressure~\cite{liu_kivi_2024,jia_saw_int4_2026}. Consequently, the optimal precision choice may differ across stages even when the same model is being served.

Taken together, these dependencies imply that stage placement cannot be determined independently from precision selection. Hardware assignment and numerical representation become part of the same deployment decision.

\subsection{Best Practices}

\textbf{BP-1: Map stages using achieved performance rather than theoretical capability.} Peak FLOPS, memory bandwidth, and vendor specifications provide useful context but are often poor predictors of serving performance. Placement decisions should be guided by measured throughput, latency, kernel coverage, and operational behavior under representative workloads. The relevant question is not which accelerator is theoretically faster, but which accelerator delivers the best performance for the stage being served.

\textbf{BP-2: Treat precision support as an executable capability.} A precision format becomes operationally meaningful only when a complete execution path exists. Kernel availability, runtime integration, quantization tooling, and deployment stability all determine whether a format is usable in practice. Placement decisions should therefore evaluate precision support at the system level rather than the hardware-specification level.

\textbf{BP-3: Size stage pools according to workload pressure.} The optimal ratio between prefill and decode resources depends on workload characteristics rather than model size alone~\cite{li_flowkv_2025}. Workloads dominated by long prompts tend to increase prefill demand, while workloads with long generations or large numbers of active sessions increase decode pressure. Resource allocation should follow observed stage pressure rather than static cluster partitions.

\begin{figure}[t]
\centering
\includegraphics[width=0.6\linewidth]{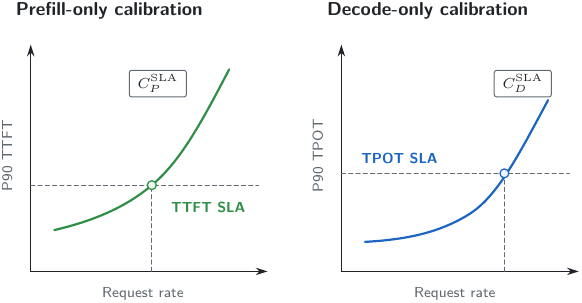}
\captionsetup{labelfont={color=black},textfont={color=black}}
\caption{Stage-isolated SLA calibration for PD pool sizing. Prefill-only and decode-only sweeps identify the request-rate knee under the relevant TTFT and TPOT objectives; these stage capacities then feed the workload-specific pool-sizing rule.}
\label{fig:pd-sla-pool-sizing}
\end{figure}

Figure~\ref{fig:pd-sla-pool-sizing} schematizes the calibration step behind this pool-sizing procedure. The deployment first profiles the prefill-only and decode-only configurations separately to find their SLA request-rate knees, denoted $C_P^{\mathrm{SLA}}$ under the TTFT objective and $C_D^{\mathrm{SLA}}$ under the TPOT objective. The PD pools are then sized so that aggregate stage capacity matches the target workload mix, for example by choosing $N_P$ and $N_D$ such that $N_P C_P^{\mathrm{SLA}} \approx N_D C_D^{\mathrm{SLA}}$.

These practices make compute placement an admission decision rather than a static hardware assignment. A stage should be mapped to an accelerator only when achieved stage performance, executable precision support, and workload-driven pool sizing point to a viable serving path; otherwise the deployment should reprofile, fall back, or colocate stages instead of relying on nominal hardware capability.
\par

\section{Making Runtime KV State Portable}
\label{sec:representation}

\subsection{Why This Decision Exists}

PD disaggregation requires that the state produced by prefill can be consumed by decode. Conceptually, the system behaves as a producer–consumer pipeline where the intermediate output (KV state) must be transferred across a stage boundary. In homogeneous deployments, this requirement is largely implicit. Both stages run within the same runtime environment and therefore share identical assumptions about tensor layout, numerical representation, and memory organization. As a result, the handoff is effectively transparent to the operator.

In heterogeneous deployments, these shared assumptions no longer hold. Prefill and decode may run on different accelerators, rely on different precision policies, use different kernel implementations, or communicate through different transport mechanisms. Under these conditions, moving raw bytes is not sufficient: the receiving stage must also interpret the state under a compatible semantic contract.

This distinction matters because transport systems typically operate at the level of buffers, descriptors, and memory regions rather than tensor semantics~\cite{mooncake_tent_docs_2026}. While some runtime connectors may provide lightweight compatibility checks, these checks are not guaranteed by the transport layer itself and must be explicitly enforced at the boundary.

As a result, heterogeneous deployments introduce a new class of failure: the transfer may succeed at the communication level, while the resulting state is unusable or inefficient at the runtime level. We therefore define the core requirement of heterogeneous PD systems as \textbf{KV state portability}: decode must be able to consume the transferred state either directly or through an explicitly validated transformation.

\subsection{Critical Dependencies}

Two dependencies consistently determine whether Runtime KV State can be transferred efficiently and consumed correctly.

\textbf{T4: Precision $\leftrightarrow$ Interconnect.}  Numerical representation directly affects what crosses the boundary. High-precision KV state is straightforward to transfer but increases bandwidth consumption and transfer latency. Lower-precision representations reduce communication volume but require both sides to agree on quantization metadata, scaling behavior, and conversion semantics~\cite{liu_kvserve_2026,yang_spectrumkv_2026}. The transfer format therefore becomes a boundary compatibility requirement rather than an implementation detail.

\textbf{T5: Interconnect $\leftrightarrow$ KV Residency.}  The cost of moving Runtime KV State is determined both by the communication link and by the placement of the state prior to and after transfer. GPU-to-GPU communication, host-staged transfers, remote-memory access, and storage-backed retrieval expose different latency and bandwidth characteristics~\cite{gopal_harvest_2026,deng_hgca_2025}. Consequently, residency and communication path must be considered jointly when evaluating transfer strategies.

These dependencies imply that KV portability is not determined solely by the transport layer. Instead, it emerges from the interaction among representation, communication path, and storage placement. The portability should be treated as a boundary contract that separates two categories of compatibility checks. The first concerns non-transformable invariants, including model and adapter identity, token range, position state, and runtime-schema compatibility. Any violation of these invariants requires rerouting or recomputation rather than transfer.

The second category covers transformable differences, such as layout, partitioning, and numerical representation. These differences can be handled through an explicit transformation plan, which may include repacking, resharding, or numerical conversion. Direct consumption is possible only when all invariants hold and no transformation is required.

\begin{figure}[t]
\centering
\includegraphics[width=0.98\linewidth]{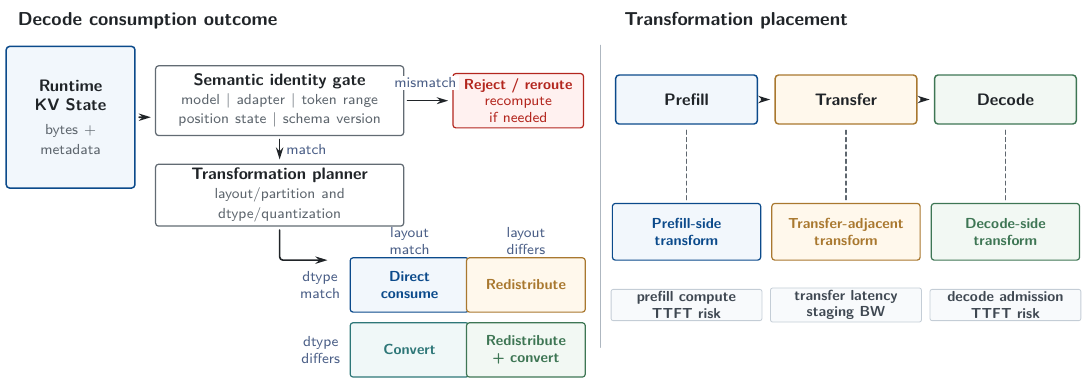}
\caption{Runtime KV State portability at the heterogeneous PD boundary. The outcome map treats semantic identity as a hard gate and compiles transformable layout, partitioning, and numerical differences into a possibly composite transformation plan. The placement map shows whether that plan consumes prefill, transfer, or decode-side first-token resources.}
\label{fig:runtime-kv-portability}
\end{figure}

The placement of this transformation determines how resources are consumed and how the first-token path is affected. Producer-side transformation can reduce boundary traffic by emitting a more compact representation, but it consumes prefill resources and may increase TTFT. Transport-adjacent transformation decouples conversion from core accelerator execution, but may introduce staging overhead and host-side bandwidth pressure. Consumer-side transformation simplifies the producer path, but delays decode admission and typically increases TTFT; it affects TPOT only when conversion is incremental or overlaps with decoding. Figure~\ref{fig:runtime-kv-portability} summarizes this consumption-and-placement model.

\subsection{Best Practices}

\textbf{BP-4: Treat Runtime KV State as a versioned boundary object.} The PD boundary should expose a well-defined Runtime KV State rather than implementation-specific tensors. Besides KV data itself, the object should carry sufficient metadata to describe representation, layout, token range, and interpretation rules. Making these assumptions explicit reduces hidden coupling between producer and consumer runtimes and simplifies future evolution of the serving stack.

\textbf{BP-5: Separate representation compatibility from byte transport.} The boundary layer should validate representation compatibility and select any required transformation, while the transport layer should move bytes between named endpoints. Conflating these responsibilities makes heterogeneous deployments fragile because correctness depends on assumptions invisible to the communication layer. Keeping them separate allows transfer mechanisms and runtime implementations to evolve independently.

\textbf{BP-6: Optimize transferred representation, not only transferred bytes.} Reducing transfer volume is often beneficial, but communication efficiency alone does not determine end-to-end performance. A representation that minimizes bandwidth consumption may still impose significant translation overhead at the receiving stage. Practical deployments should evaluate the combined cost of transfer, conversion, validation, and consumption rather than optimizing any single component in isolation. For example, W4A8 may reduce memory-bandwidth traffic, but later de-quantization can offset that saving; the fixed-load results in Table~\ref{tab:h200-precision-fixed} show why low-bit paths must be judged by end-to-end impact.

These practices make Runtime KV State portability an explicit boundary decision. The system should name the state being handed off, validate the representation assumptions needed for decode-side consumption, keep those checks separate from byte transport, and choose the transferred representation by its combined transfer, conversion, validation, and consumption cost.

\section{Managing Runtime KV State Beyond the Handoff}
\label{sec:handoff}

\subsection{Why This Decision Exists}
Successfully transferring Runtime KV State across the PD boundary does not complete the handoff. Even after the state arrives at the decode stage, the system must continue managing its lifetime, placement, and lifecycle responsibilities~\cite{vllm_nixl_lease_docs_2026,sglang_pd_docs_2026}. In homogeneous deployments, these responsibilities are typically hidden within a single runtime. Heterogeneous deployments expose them explicitly, because different components may participate in producing, storing, transferring, and consuming the same state.

\begin{table}[!b]
\centering
\caption{KV ownership, release, and cleanup responsibilities in selected vLLM
and SGLang PD handoff paths.}
\label{tab:s6-handoff}
{\small
\setlength{\tabcolsep}{3pt}
\begin{tabular}{@{}L{0.16\linewidth}L{0.20\linewidth}L{0.20\linewidth}L{0.18\linewidth}L{0.18\linewidth}@{}}
\toprule
Runtime path & Movement trigger & Capacity ownership & Normal release & Cleanup / recovery \\
\midrule
vLLM NIXL pull / lease~\cite{vllm_nixl_lease_docs_2026,vllm_nixl_source_2026} & Decode posts \texttt{READ} after prefill returns block
coordinates. & Decode allocates destination; prefill pins source under renewable
lease. & Read-completion notification releases prefill blocks. &
Heartbeat loss stops renewal; lease expiry reclaims source blocks after crash or notification loss. \\
\addlinespace
vLLM NIXL push~\cite{vllm_nixl_push_docs_2026,vllm_nixl_lease_docs_2026,vllm_nixl_source_2026} & Decode sends \texttt{PUSH\_REG}; prefill matches and writes
into decode memory. & Decode preallocates destination; prefill holds source and
match state. & WRITE completion releases the P-side lease and marks the D-side receive complete. &
The D-side watchdog drops stale registrations, while request abort reclaims destination-side state; P-side lease expiry reclaims source blocks. \\
\addlinespace
SGLang PD path~\cite{sglang_pd_docs_2026,sglang_pd_source_2026} & Decode preallocates destinations and sends metadata; prefill
transfers via Mooncake or NIXL. & Decode owns preallocation and admission
gating; prefill holds sender and cache-lock state. & Transfer completion makes received KV eligible for decode; prefill releases source KV, and sender/receiver state is cleared. & Bootstrap and waiting timeouts; abort
and failed-decode release clear uncommitted capacity. \\
\bottomrule
\end{tabular}
}
\end{table}

The portability check only determines whether a valid transfer path exists between producer and consumer. It does not specify which component pins the source state, which entity accounts for destination capacity, which event marks transfer completion, or when release and cleanup become safe.
This separation raises a more fundamental question: lifecycle responsibilities must be assigned at each handoff point, rather than assuming a single end-to-end owner of the state.
The answer directly impacts resource utilization, admission control, failure handling, and system scalability. If release responsibility remains with the producer, resources may be retained longer than necessary. If capacity accounting is delegated immediately to the consumer, the receiving stage must pre-reserve capacity before consumption begins. When multiple storage tiers participate in the serving path, lifecycle responsibilities may be distributed across several components.

As deployments introduce KV offloading, shared caches, remote memory, and storage-backed retrieval~\cite{liu_cachegen_2023,strati_dejavu_2024}, the lifecycle of Runtime KV State extends beyond a simple producer–consumer interaction and spans source pinning, destination reservation, transfer, commit, release, and cleanup. Managing this lifecycle therefore becomes a first-class systems concern.
Table~\ref{tab:s6-handoff} distills three concrete PD handoff paths from vLLM and SGLang, at the pinned revisions inspected in this paper, into the lifecycle responsibilities they expose: movement trigger, capacity ownership, release, and cleanup/recovery.

\subsection{Critical Dependencies}

Two dependencies consistently shape ownership and lifecycle decisions.

\textbf{T6: Stage Pressure $\leftrightarrow$ KV Residency.} The live and retained Runtime KV State footprint depends heavily on workload conditions. Long prompts, long generations, high concurrency, and prefix reuse all increase pressure on KV capacity. As pressure changes, the preferred residency strategy may shift among GPU memory, host memory, shared caches, and colder storage tiers. Consequently, lifecycle management and residency management cannot be separated.
Moreover, residency decisions implicitly define future access paths. A state object that resides in local GPU memory can be consumed immediately, while a state object placed in remote memory or persistent storage requires retrieval before use. As a result, ownership decisions influence not only capacity utilization but also future communication costs and latency behavior.

These dependencies imply that Runtime KV State should not be viewed as a static cache entry. It is a moving system resource whose value and accessibility change throughout its lifetime.

\subsection{Best Practices}

\textbf{BP-7: Make lifecycle responsibilities explicit.} Lifecycle responsibility should change through well-defined handoff points rather than implicit runtime assumptions. The system should be able to determine which component is responsible for physical holding, capacity accounting, transfer completion, release, and cleanup at any moment. Explicit lifecycle responsibilities simplify resource accounting, failure recovery, and operational debugging.

\textbf{BP-8: Separate retention policy from residency policy.} The decision to retain Runtime KV State is conceptually different from the decision of where to place it. A deployment may choose to retain state for future reuse while continuously migrating it across storage tiers. Keeping these policies independent allows systems to adapt residency according to workload pressure without changing application-level reuse behavior.

\begin{figure}[t]
\centering
\includegraphics[width=0.8\linewidth]{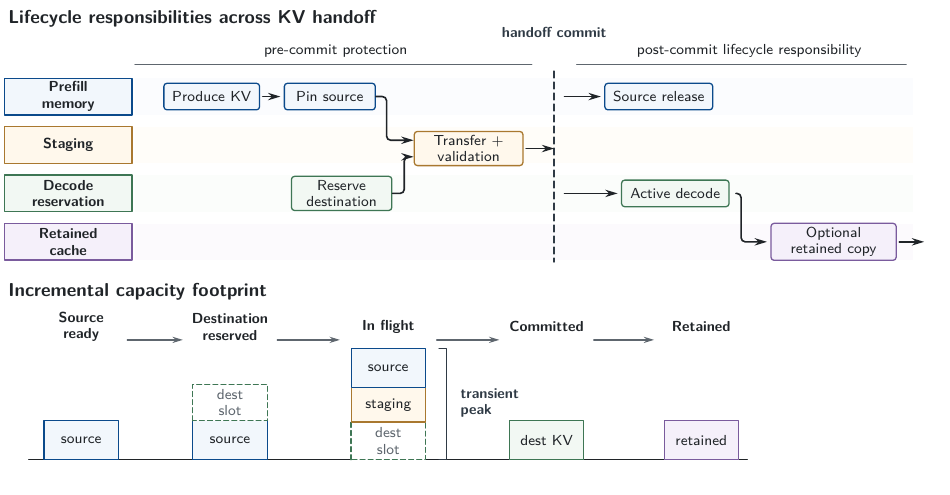}
\caption{KV handoff commit and transient capacity. The commit point shifts lifecycle responsibility and exposes the transient footprint that admission control must budget.}
\label{fig:handoff-ownership}
\end{figure}

\textbf{BP-9: Treat KV capacity as a shared serving resource.} In large-scale deployments, KV capacity becomes a scheduling resource comparable to compute and network bandwidth. Admission decisions, cache retention, and offloading policies should therefore be evaluated together rather than independently. Optimizing any single component in isolation often shifts pressure elsewhere in the system.

These practices treat Runtime KV State as a lifecycle-managed serving resource: retention decides whether state remains useful, residency decides where it is held, and lifecycle responsibility decides which component accounts for capacity, commits the handoff, releases source state, and cleans up failures. Figure~\ref{fig:handoff-ownership} illustrates the resulting commit point: before commit, source state, staging buffers, and destination reservations may coexist, so admission control must budget the transient footprint together with active-decode KV and retained cache.
\par

\section{Empirical Evidence}

We present empirical evidence supporting the three boundary decisions described in Sections~\ref{sec:placement}--\ref{sec:handoff}. We use one production deployment and a set of controlled single-node measurements. The deployment case demonstrates how all decisions co-exist in a real heterogeneous PD system, while the controlled experiments isolate individual mechanisms under fixed execution conditions.

\subsection{Production Deployment (CPHD-GLM5.1)}

We first examine an observed CPHD-GLM5.1 deployment, where CPHD refers to a heterogeneous PD setup with prefill on MetaX C600 and decode on NVIDIA Hopper accelerator, instantiating a full PD serving pipeline in production. As shown in Table~\ref{tab:cphd-glm51-config}, prefill and decode are assigned to different accelerator pools (MetaX C600 and Hopper), resulting in a fixed stage split and stage-specific parallelism configuration.

\begin{table}[H]
\centering
\caption{Heterogeneous PD deployment configuration of CPHD-GLM5.1.}
\label{tab:cphd-glm51-config}
{\small
\setlength{\tabcolsep}{4pt}
\begin{tabular}{@{}llcL{0.25\linewidth}L{0.22\linewidth}@{}}
\toprule
Stage & Accelerator & Units & Parallelism & Precision path \\
\midrule
Prefill & MetaX C600 & $4{\times}8$ & TP4-CP4-PP4 & INT8 (W8A8) \\
Decode & NVIDIA Hopper & $2{\times}8$ & TP16-DP16-EP16; DP attention & FP8 \\
\bottomrule
\end{tabular}
}
\end{table}

This deployment simultaneously determines three coupled decisions: (i) compute placement across heterogeneous accelerators, (ii) the precision path available to each stage (INT8 for prefill and FP8 for decode), and (iii) the effective P:D workload partitioning. Table~\ref{tab:cphd-glm51-results} shows that under a high prefix-cache workload, the system achieves stable throughput and satisfies end-to-end latency constraints over a prolonged execution window.

\begin{table}[H]
\centering
\caption{Serving workload, throughput, and SLA of CPHD-GLM5.1.}
\label{tab:cphd-glm51-results}
{\small
\setlength{\tabcolsep}{5pt}
\begin{tabular}{@{}ll@{}}
\toprule
Metric & Value \\
\midrule
Input length & 64K \\
Output length & 512 \\
Prefix-cache hit rate (\%) & 90 \\
Arrival rate (requests/s) & 1.8 \\
Service request throughput (requests/s) & 1.62 \\
Prefill throughput (tokens/s) & 107.7K \\
Decode throughput (tokens/s) & 827 \\
TTFT p50 (s) & $\leq$ 5 \\
TPOT p90 (ms) & $\leq$ 30 \\
Steady run (h) & 5 \\
\bottomrule
\end{tabular}
}
\end{table}

Table~\ref{tab:cphd-glm51-quality} further reports that model quality remains within acceptable deviation bounds compared to official references, indicating that heterogeneous execution does not introduce measurable degradation under the tested configuration.

\begin{table}[H]
\centering
\caption{Evaluation on benchmarks.}
\label{tab:cphd-glm51-quality}
{\small
\setlength{\tabcolsep}{4pt}
\begin{tabular}{@{}lccc@{}}
\toprule
Benchmark & Official ref. & Observed & Abs. gap \\
\midrule
AIME 25 & 93--97 & 93 & 0.0 \\
AIME 26 & 95.3 & 93 & 2.3 \\
SWE-Bench Verified & 76.4 & 73.4 & 3.0 \\
\bottomrule
\end{tabular}
}
\end{table}

\subsection{Compute Placement Validation}

We next validate the compute placement decision using controlled SLA-style measurements on a single-node Hopper setup. Table~\ref{tab:h200-sla-breakpoints} varies stage split and KV representation under identical workload conditions.
\textbf{}
The results show that compute placement is highly sensitive to both workload pressure and KV format. For BF16 KV, changing the P:D split from 6P2D to 4P4D reduces the maximum sustainable request rate $r^*$ from 0.2 to 0.1. At a fixed 4P4D configuration, changing KV representation from BF16 to FP8 increases $r^*$ from 0.1 to 1.0, demonstrating that compute placement cannot be separated from KV representation.

\begin{table}[H]
\centering
\caption{Single-node controlled SLA-style breakpoints for Qwen3-32B SGLang PD (NIXL, BF16 weights, 4K-input/512-output). $r^*$ is the highest offered request rate satisfying p90 TTFT $<10$s and p99 TPOT $<25$ms.}
\label{tab:h200-sla-breakpoints}
{\small
\begin{tabular}{@{}llrrrrrrr@{}}
\toprule
Topology & KV dtype & $r^*$ & Req/s & \multicolumn{3}{c}{TTFT (s)} & TPOT p99 (ms) & ITL p99 (ms) \\
\cmidrule(lr){5-7}
 & & & & p50 & p90 & p99 & & \\
\midrule
4P4D & \texttt{bf16} & 0.1 & 0.115 & 3.35 & 6.29 & 10.88 & 20.07 & 20.41 \\
6P2D & \texttt{bf16} & 0.2 & 0.226 & 4.36 & 9.89 & 20.25 & 20.78 & 21.49 \\
4P4D & \texttt{fp8\_e4m3} & 1.0 & 0.888 & 3.72 & 8.75 & 13.80 & 21.72 & 22.80 \\
\bottomrule
\end{tabular}
}
\end{table}

\subsection{KV Representation Impact}

We further isolate the effect of KV representation under fixed compute placement and workload conditions. Table~\ref{tab:h200-sla-breakpoints} and Table~\ref{tab:h200-precision-fixed} show that KV dtype affects both throughput and latency behavior.

Under identical 4P4D placement, FP8 KV significantly increases throughput compared to BF16, while shifting tail latency behavior. These effects are asymmetric across metrics, indicating that KV representation acts as a system-level constraint.

\begin{table}[H]
\centering
\caption{Single-node controlled stressed fixed-load precision-path comparison for Qwen3-32B SGLang PD (NIXL, 4P4D, 4K-input/512-output, offered request rate $r=1.0$). KV dtype is the runtime default for each model path.}
\label{tab:h200-precision-fixed}
{\small
\begin{tabular}{@{}llrrrrr@{}}
\toprule
Model path & KV dtype & Req/s & Tok/s & TTFT p99 (s) & TPOT p99 (ms) & ITL p99 (ms) \\
\midrule
BF16 & \texttt{bf16} & 0.673 & 3.10K & 32.61 & 21.02 & 21.89 \\
FP8 & \texttt{bf16} & 0.701 & 3.23K & 31.79 & 13.74 & 14.39 \\
AWQ INT4 & \texttt{fp16} & 0.761 & 3.51K & 36.55 & 10.24 & 10.54 \\
\bottomrule
\end{tabular}
}
\end{table}

\subsection{Precision-Path Tradeoffs}

We finally evaluate precision-path effects under fixed placement and workload conditions. The results show that FP8 and AWQ both improve decode-side efficiency relative to BF16, but introduce different tradeoffs in prefill latency (TTFT). AWQ improves TPOT p99 but increases TTFT p99, demonstrating asymmetric effects across stages.
These results reinforce the coupling between precision policy and stage-specific workload pressure.

\section{Discussion}
\label{sec:open-coupling-interconnect-stage}

Most coupling cells analyzed above yield concrete boundary practices. Two cells, however, are better treated as open system questions because their resolution depends on both the communication substrate and the degree of control available to the system over accelerator and network provisioning. Figure~\ref{fig:couplings} highlights these open cells (O1 \& O2). We discuss them below: cross-accelerator KV movement and stage--interconnect co-design.

\textbf{Cross-accelerator KV movement depends on heterogeneous communication stacks.} 

Systems such as FlagCX illustrate how this path is typically realized in practice: they wrap vendor-specific collectives behind a unified adaptor layer, exposing device-buffer transport as part of the runtime communication stack~\cite{flagcx_project_2026}. However, correctness and performance are ultimately constrained by the alignment of accelerators, runtime libraries, memory registration mechanisms, and network backends, which remain outside the scope of a purely portable abstraction. This makes KV movement a property of the end-to-end communication stack rather than a transport-layer feature alone.

\textbf{Stage pressure interacts with interconnect as a provisioning variable.} Stage pressure not only affects request routing, prefill chunking, cache placement, and the prefill-to-decode capacity ratio, but also determines the effective demand placed on the underlying network fabric. In deployments where the operator controls hardware provisioning, interconnect bandwidth and topology become co-design variables alongside compute placement. In this regime, KV traffic is not merely scheduled over a fixed network; it can influence how many network resources (e.g., NIC bandwidth, cross-rack links, or cluster-level connectivity) are provisioned to match the expected PD workload mix. Systems such as Prefill-as-a-Service illustrate the low-transfer regime, where KV-efficient architectures reduce communication volume sufficiently to enable cross-cluster execution, with performance governed jointly by stage placement and network capacity~\cite{qin_prefill_as_a_service_2026}. The remaining open question is when interconnect should be treated as a fixed constraint handled by software, and when it becomes a first-order design parameter coupled with PD stage allocation.

\section{Related Work}

We structure related work by the system mechanisms they address, and position each line of work within our design space formulation.

\textbf{PD disaggregation and phase splitting.} Separating prefill and decode
into independently provisioned pools is an established
technique~\cite{zhong_distserve_2024,patel_splitwise_2023,hu_tetriinfer_2024},
built on memory-managed serving runtimes~\cite{kwon_pagedattention_2023} and
extended toward specialized per-stage hardware~\cite{zhang_spad_2025} and the
control of prefill-decode contention~\cite{lin_heterogeneous_workloads_2026}.
This work establishes that the two stages differ enough to plan separately; we
take the next step of treating the accelerator, precision, interconnect, and
residency choices that the split exposes as coupled axes at the boundary.

\textbf{Heterogeneous and multi-vendor serving.} Recent systems run prefill and
decode on different accelerators: a multi-vendor GPU PD system with a
cross-vendor transmission layer~\cite{chen_disaggregated_multivendor_2025} and a
large heterogeneous-NPU deployment that places decode on the
higher-interconnect parts of the fleet~\cite{xiao_xdeepserve_2025}. These are
point solutions on particular accelerator and interconnect configurations; our
contribution is the coupling analysis that tells an operator which of those
choices can be made independently and which must be co-designed.

\textbf{Scheduling under workload heterogeneity.} Workload-aware scheduling
balances prefill and decode load and routes KV
transfers~\cite{hu_tetriinfer_2024,li_flowkv_2025}. Workload enters our framework
through stage pressure, and we use it to explain why the P:D pool ratio and
handoff timing shift with the workload regime.

\textbf{Quantization and precision policy.} Low-bit weight and activation
quantization~\cite{lin_awq_2024,lin_qserve_2024,lu_mixquant_2026} and KV-cache
quantization~\cite{liu_kivi_2024,jia_saw_int4_2026} reduce compute and memory
pressure. Phase-aware precision allocation assigns different bitwidths to
prefill and decode based on their distinct
bottlenecks~\cite{chen_pmpd_2024,zhang_qqq_2024}. These mechanisms underlie the
precision axis; the coupling analysis adds that a bitwidth label becomes a
deployable policy only when an executable backend path and a named runtime role
exist at the PD boundary.

\textbf{KV transfer, compression, and residency.} A growing body of work moves
and stores KV state across tiers: KV-centric disaggregated serving and its
transfer runtime~\cite{qin_mooncake_2024,mooncake_tent_docs_2026}, cache layers
and streaming~\cite{liu_lmcache_2025,strati_dejavu_2024,liu_cachegen_2023},
service-aware compression for disaggregated
transfer~\cite{liu_kvserve_2026,yang_spectrumkv_2026}, peer-to-peer and GPU-CPU
residency~\cite{gopal_harvest_2026,deng_hgca_2025}, and RDMA point-to-point
transport~\cite{licker_fabriclib_2025}. These mechanisms define the transfer and
storage options behind two boundary decisions in our framework: the runtime KV
format that crosses the PD boundary, and the online-path eligibility of each
residency tier.

\textbf{Network connectivity and interconnect provisioning.} Network-level work
targets the connectivity substrate beneath the PD boundary: portable
point-to-point RDMA for LLM systems~\cite{licker_fabriclib_2025}, cross-chip
communication libraries for heterogeneous accelerators~\cite{flagcx_project_2026},
datacenter network topologies for large AI clusters~\cite{liao_ubmesh_2025}, and
cross-datacenter prefill enabled by smaller transferred KV
state~\cite{qin_prefill_as_a_service_2026}. These efforts explain why
interconnect is not only an implementation detail in our framework: it determines
which residency tiers and stage placements remain online-eligible under a
particular deployment topology.
\par

\section{Conclusion}

Heterogeneous PD inference is becoming a normal serving mode, but its risk lies
in treating accelerator, precision, interconnect, stage pressure, and KV
residency as independent knobs. This paper maps these five axes and demystifies how their strongest dependencies reduce to three boundary decisions: where
computation should run, how Runtime KV State is represented and consumed, and
who owns that state through reservation, transfer, release, and recovery. The
resulting best practices give operators a practical way to adopt heterogeneity:
profile stage placement by achieved performance, validate executable precision
and KV-representation paths at the boundary, and manage KV capacity as a
lifecycle resource across prefill and decode.
\par

\bibliographystyle{unsrt}
\bibliography{references}

\end{document}